\documentstyle[amssymb,aps]{revtex}

\begin{document}
\author{O. G. Balev}
\address{Departmento de F\'{i}sica, Universidade Federal de S\~{a}o Carlos,\\
13565-905, S\~{a}o Carlos, Brazil and Institute of Semiconductor Physics,\\
National Academy of Sciences, 45 Pr. Nauky, Kiev, 252650, Ukraine}
\author{F. T. Vasko}
\address{Departmento de F\'{i}sica, Universidade Federal de Mato Grosso do Sul,\\
79070-900, Campo Grande, Brazil and Institute of Semiconductor Physics,\\
National Academy of Sciences, 45 Pr. Nauky, Kiev, 252650, Ukraine }
\author{Fl\'{a}vio Aristone}
\address{Departmento de F\'{i}sica, Universidade Federal de Mato Grosso do Sul,\\
79070-900 Campo Grande, Brazil}
\author{Nelson Studart}
\address{Departmento de F\'{i}sica, Universidade Federal de S\~{a}o Carlos,\\
13565-905, S\~{a}o Carlos, Brazil \\
}
\date{February 28, 2000}
\title{Two-subband electron transport in nonideal quantum wells }
\maketitle

\begin{abstract}
Electron transport in nonideal quantum wells (QW) with large-scale
variations of energy levels is studied when two subbands are occupied.
Although the mean fluctuations of these two levels are screened by the
in-plane redistribution of electrons, the energies of both levels remain
nonuniform over the plane. The effect of random inhomogeneities on the
classical transport is studied within the framework of a local response
approach for weak disorder. Both short-range and small-angle scattering
mechanisms are considered. Magnetotransport characteristics and the
modulation of the effective conductivity by transverse voltage are evaluated
for different kinds of confinement potentials (hard wall QW, parabolic QW,
and stepped QW).

PACS\ \ 73.20.Dx; 73.40.-c; 72.10.Fk
\end{abstract}

\section{Introduction}

The study of transport properties of semiconductor structures with many
occupied electronic subbands has attracted scientific interest for a long
time, concerning especially with two-subband occupancy and screening effects
in the scattering processes \cite{1,2,3,4,5,6,7,8,9,10,11}. Moreover, until
recently these studies were made in inversion layers of Si-MOS
(metal-oxide-semiconductor) structures \cite{1,2} in GaAs/Al$_{x}$Ga$_{1-x}$%
As selectively doped heterojunctions with two-subband occupied \cite{3,4,5},
where large-scale fluctuations of layers have small effect on the position
of the energy levels and transport properties. Recently, new types of
quantum wells (QW), e.g., modulation-doped QW \cite{7}, wide parabolic QW 
\cite{8,9,11}, and stepped QW \cite{10} semiconductor structures with two
occupied subbands, have been grown by tailoring the conduction-band edge of
III-IV semiconductors. It was shown that these novel heterostuctures with
double subband occupancy exhibit a much stronger effect of large-scale
disorder on transport properties than the previously studied one-subband
systems. For the new systems, the screening is effective only on the average
of the external changes of the bottoms of the two occupied subbands, which
are different from each other, and caused by smooth variations of QW
parameters on the scale of the Bohr radius as illustrated in Fig. 1a. Only
essential unscreened large-scale fluctuations of the both occupied QW levels
are present in these systems.

This paper addresses the effect of such non-screened variations of both
occupied energy levels on the electronic spectrum and the classical
magnetotransport by employing different confinement models: hard wall QW
(Fig. 1b), parabolic QW (Fig. 1c), and stepped QW (Fig. 1d). Short-range and
small-angle elastic scattering mechanisms, within a local response
approximation, are included in the linearized kinetic equation. We obtain
effective magnetotransport coefficients, which are evaluated from the
components of the local conductivity tensor averaged over large-scale
inhomogeneities, up to second order in the disorder contributions to the
conductivity\cite{12}, for zero temperature. In addition, we calculate the
modulation of the effective conductivity by the transverse voltage. We take
into account the difference between intra-subband momentum scattering
frequencies and the inter-subband contributions. We investigate also the
spatial variations of these frequencies. We find an essential broadening of
the peak of intersubband transitions due to the large-scale inhomogeneities
and we provide numerical estimates of magnetotransport coefficients for the
considered models.

The organization of the paper is as follows. In Sec. II the energy spectrum,
which takes into account non-screened variations of levels, and the general
conductivity tensor within the linear response approach, are obtained. From
the transition probability for a random potential model, we calculate, in
Sec. III, the scattering frequencies and magnetotransport coefficients
considering short-range and small-angle scattering mechanisms and after
averaging the conductivity tensor over in-plane large-scale inhomogeneities.
In Sec. IV, we discuss the assumptions, on which the work described here is
explicitly dependent, and we present our concluding remarks.

\section{General equations}

We discuss now our model for the energy levels fluctuations due to random
inhomogeneities and we obtain the general expressions of the conductivity
tensor based on the local response approach for nonideal QWs with two
occupied subbands.

\subsection{Electron energy spectrum}

Electron energy levels in a QW with slowly varying width can be described in
terms of ${\bf r}=\{x,y\}$ along with the in-plane kinetic energy given by
the parabolic dispersion law $\varepsilon _{p}=p^{2}/2m$ with effective mass 
$m$. For instance, the influence of nonideal heterointerfaces in the
hard-wall QW, formed by large band offsets, can be studied by assuming that
the QW widths, $d_{{\bf r}}$, are randomly varied. Then, the few lowest
energy levels $s=1,2,...$ are given by

\begin{equation}
\varepsilon _{s{\bf r}}=\frac{(s\pi \hbar /d_{{\bf r}})^{2}}{2m}\simeq
\varepsilon _{s}\left( 1-2\frac{\delta d_{{\bf r}}}{d}\right) ,  \label{1}
\end{equation}
where $\varepsilon _{s}=s^{2}(\pi \hbar /d)^{2}/2m$, $d$ is the mean width
of the QW, and $\delta d_{{\bf r}}$ is its random fluctuation due to
heterointerface roughness. The energy of an electron in the $s$-th subband
is given as $E_{p{\bf r}}^{(s)}=p^{2}/2m+\varepsilon _{s{\bf r}}+v_{s{\bf r}%
} $, where $v_{s{\bf r}}=\int dz\left| \varphi _{sz}\right| ^{2}V_{{\bf r}z}$
is the screening potential averaged over the $s$-th subband charge
distribution $\left| \varphi _{sz}\right| ^{2}$ along the $z$ direction. The
potential $V_{{\bf r}z}$ is determined by the Poisson's equation as

\begin{equation}
\nabla ^{2}V_{{\bf r}z}=-\frac{4\pi e^{2}}{\epsilon }(n_{{\bf r}z}-\langle
n_{{\bf r}z}\rangle ),~~~n_{{\bf r}z}=\sum_{s=1,2}n_{s{\bf r}}\left| \varphi
_{sz}\right| ^{2},~~~  \label{2}
\end{equation}
where $n_{s{\bf r}}=\rho _{2D}(\varepsilon _{F}-\varepsilon _{s{\bf r}}-v_{s%
{\bf r}})$ is the electron density in the $s$-th subband, $\rho _{2D}=m/\pi
\hbar ^{2}$ is the 2D density of states, $\varepsilon _{F}$ the Fermi
energy, $\langle n_{{\bf r}z}\rangle $ is the electron density averaged over
the random inhomogeneities, and $\epsilon $ is the dielectric constant,
assumed independent of $z$. Hereafter, we consider only the case of two
occupied subbands.

The solution of Eq. (\ref{2}) is given as

\begin{equation}
V_{{\bf q}z}=\frac{2\pi e^{2}}{\epsilon q}\int dz^{\prime }e^{-q|z-z^{\prime
}|}\delta n_{{\bf q}z^{\prime }},  \label{3}
\end{equation}
where $\delta n_{{\bf q}z}$ is the 2D Fourier transform of the nonuniform
part of the electron density

\begin{equation}
\delta n_{{\bf r}z}=\sum_{s=1,2}\left| \varphi _{sz}\right| ^{2}\delta n_{s%
{\bf r}}=-\rho _{2D}\sum_{s}\left| \varphi _{sz}\right| ^{2}(\delta
\varepsilon _{s{\bf r}}+v_{s{\bf r}}).  \label{4}
\end{equation}
Substituting Eq. (\ref{4}) in Eq. (\ref{3}), we are led to the system of two
linear inhomogeneous equations for the Fourier transforms of screening
potentials $v_{s{\bf q}}$ for $s=1$ and $s=2$ subbands:

\begin{equation}
v_{s{\bf q}}=-\frac{2\pi e^{2}}{\epsilon q}\rho _{2D}\sum_{s^{\prime }}\int
dz\varphi _{sz}^{2}\int dz^{\prime }e^{-q|z-z^{\prime }|}\left| \varphi
_{s^{\prime }z^{\prime }}\right| ^{2}(\delta \varepsilon _{s^{\prime }{\bf q}%
}+v_{s^{\prime }{\bf q}}).  \label{5}
\end{equation}
Notice that inhomogeneity of Eq. (\ref{5}) comes from terms $\propto \delta
\varepsilon _{s^{\prime }{\bf q}}$. Hence the solutions are expressed in
terms of the variations of the occupied levels, introduced by Eq. (\ref{1}).
Assuming large-scale variations, $\ell _{c}\gg d$, it follows from Eq. (\ref
{5}) that only the Fourier components with $q\alt\ell _{c}^{-1}\ll d^{-1}$
are essential. Then $\exp (-q|z-z^{\prime }|)$ in Eq. (\ref{5}) can be well
approximated by the unity and we have

\begin{equation}
v_{s{\bf q}}=-\frac{4}{qa_{B}}\sum_{s^{\prime }}(\delta \varepsilon
_{s^{\prime }{\bf q}}+v_{s^{\prime }{\bf q}}),  \label{6}
\end{equation}
where $a_{B}$ is the Bohr radius. Here the right-hand side is independent of 
$s$, so $v_{1,2{\bf q}}\equiv v_{{\bf q}}$. Further, for $qa_{B}/8\ll 1,$
from Eq. (\ref{6}), it follows that

\begin{equation}
v_{{\bf r}}\simeq -\frac{1}{2}\sum_{s=1,2}\delta \varepsilon _{s{\bf r}},
\label{7}
\end{equation}
i.e., screening is relevant only on the average of the external variations $%
\delta \varepsilon _{1{\bf r}}$ and $\delta \varepsilon _{2{\bf r}}$. As a
result, the electron spectrum energy can be written in the form

\begin{equation}
E_{p{\bf r}}^{(s)}=\frac{p^{2}}{2m}+\varepsilon _{s}-(-1)^{s}\delta
\varepsilon _{{\bf r}},  \label{8}
\end{equation}
where $\delta \varepsilon _{{\bf r}}=(\varepsilon _{2}-\varepsilon
_{1})\delta d_{{\bf r}}/d\approx 3\varepsilon _{1}\delta d_{{\bf r}}/d$ is
the non-screened part of potential. For further numerical estimates, we
assume Gaussian correlations for the QW width fluctuations

\begin{equation}
\langle \delta \varepsilon _{{\bf r}}\delta \varepsilon _{{\bf r^{\prime }}%
}\rangle =\left( \frac{3\varepsilon _{1}}{d}\right) ^{2}\langle \delta d_{%
{\bf r}}\delta d_{{\bf r^{\prime }}}\rangle =\left( \overline{\delta
\varepsilon }\right) ^{2}w(|{\bf r}-{\bf r^{\prime }}|).  \label{9}
\end{equation}
Here $\overline{\delta \varepsilon }=3\varepsilon _{1}\overline{\delta d}/d$%
, $\overline{\delta d}$ is the typical height of the roughness interface and 
$w(x)=\exp [-(x/\ell )^{2}]$ is the Gaussian function with lateral
correlation length $\ell $. We can also write analogous expressions for the
case of parabolic (Fig. 1c) and stepped (Fig.1d) QWs but we do not give here
the pertinent formulas considering further that $\overline{\delta
\varepsilon }$ is a given quantity.

\subsection{Conductivity tensor}

Classical transport phenomena for electrons with spectrum, given by Eq. (\ref
{8}), are described by the distribution function $f_{F}(E_{p{\bf r}%
}^{(s)})+\Delta f_{{\bf pr}}^{(s)}$ within the linear response theory. Here $%
f_{F}$ is the unperturbed Fermi distribution and the field-induced
distribution is determined from the linearized kinetic equation of $s$-th
subband given by

\begin{equation}
\left( {\bf v\cdot }{\bf \nabla }_{{\bf r}}+(-1)^{s}{\bf \nabla }\delta
\varepsilon _{{\bf r}}\cdot \frac{\partial }{\partial {\bf p}}+\frac{e}{c}[%
{\bf v}\times {\bf H}]\cdot \frac{\partial }{\partial {\bf p}}\right) \Delta
f_{{\bf pr}}^{(s)}-I_{s}(\Delta f|{\bf pr})=-e{\bf E}_{{\bf r}}\cdot \frac{%
\partial f_{F}(E_{p{\bf r}}^{(s)})}{\partial {\bf p}},  \label{10}
\end{equation}
where ${\bf E}_{{\bf r}}$ is the electric field, which is zero in the
absence of a net current density ${\bf j}_{{\bf r}}$, ${\bf H}$ is the
magnetic field perpendicular to the QW's plane (${\bf v}={\bf p}/m$) and the
collision integral for the elastic scattering case is written as

\begin{equation}
I_{s}(\Delta f|{\bf pr})=\sum_{s^{\prime }{\bf p^{\prime }}}W_{{\bf %
pp^{\prime }}}^{ss^{\prime }}(\Delta f_{{\bf p^{\prime }r}}^{(s^{\prime
})}-\Delta f_{{\bf pr}}^{(s)}).  \label{11}
\end{equation}
$W_{{\bf pp^{\prime }}}^{ss^{\prime }}$ is the transition probability per
second between states $s,{\bf p}$ and $s^{\prime },{\bf p^{\prime }}$. The
total current density ${\bf j}_{{\bf r}}$ is given by the standard
expression ${\bf j}_{{\bf r}}=(2e/L^{2})\sum_{s{\bf p}}{\bf v}\Delta f_{{\bf %
pr}}^{(s)}$ and satisfies the continuity equation $\nabla \cdot {\bf j}_{%
{\bf r}}=0$.

The contributions from the convection and random fields in the left-hand
side of Eq. (\ref{10}) are estimated as $\bar{v}/\ell $ and $\overline{%
\delta \varepsilon }/(\bar{p}\ell )$ respectively. These contributions can
be discarded when $\bar{v}\bar{\tau}\ll \ell $ and $\overline{\delta
\varepsilon }\ll \bar{\varepsilon}$, where $\bar{\varepsilon}=\bar{v}\bar{p}%
/2$ is the characteristics electron energy, where we have used the
relaxation time approximation to estimate the collision integral as $\bar{%
\tau}^{-1}.$ Under these assumptions, Eq. (\ref{10}) assumes the form

\begin{equation}
\frac{e}{c}[{\bf v}\times {\bf H}]\frac{\partial }{\partial {\bf p}}\Delta
f_{{\bf pr}}^{(s)}-I_{s}(\Delta f|{\bf pr})=-e({\bf E}_{{\bf r}}\cdot {\bf v}%
)f_{F}^{^{\prime }}(E_{p{\bf r}}^{(s)}),  \label{13}
\end{equation}
where $f_{F}^{^{\prime }}(E)=\partial f_{F}(E)/\partial E$. We look for a
solution in the form

\begin{equation}
\Delta f_{{\bf pr}}^{(s)}=-ef_{F}^{^{\prime }}(E_{p{\bf r}}^{(s)})\left\{
A_{s}({\bf E}_{{\bf r}}\cdot {\bf v})+B_{s}\frac{e}{mc}([{\bf E}_{{\bf r}%
}\times {\bf H}]\cdot {\bf v})\right\} ,  \label{14}
\end{equation}
where the coefficients $A_{s}$ and $B_{s}$ depend only on $p^{2}$.
Substituting Eq. (\ref{14}) into the kinetic equation (\ref{13}), we obtain
the linear algebraic system (see Ref.\cite{11} for details)

\begin{eqnarray}
\omega _{c}^{2}B_{s}-\sum_{s^{\prime }{\bf p^{\prime }}}W_{{\bf pp^{\prime }}%
}^{ss^{\prime }}\left( A_{s^{\prime }}\frac{p^{\prime }}{p}\cos ({\bf p},%
{\bf p^{\prime }})-A_{s}\right) &=&1,  \nonumber \\
A_{s}+\sum_{s^{\prime }{\bf p^{\prime }}}W_{{\bf pp^{\prime }}}^{ss^{\prime
}}\left( B_{s^{\prime }}\frac{p^{\prime }}{p}\cos ({\bf p},{\bf p^{\prime }}%
)-B_{s}\right) &=&0.  \label{15}
\end{eqnarray}
The solution of Eq. (\ref{15}) for our two-level system is expressed through
the transport relaxation frequency of $s$-th subband given by

\begin{equation}
\nu _{s}^{tr}=\sum_{{\bf p^{\prime }}}W_{{\bf pp^{\prime }}}^{ss}\left[
1-\cos ({\bf p},{\bf p^{\prime })}\right]  \label{16}
\end{equation}
and the intersubband relaxation frequencies are written as

\begin{equation}
\nu _{ss{\bf ^{\prime }}}=\sum_{{\bf p^{\prime }}}W_{{\bf pp^{\prime }}}^{ss%
{\bf ^{\prime }}},\;\;\;\tilde{\nu}_{ss{\bf ^{\prime }}}=\sum_{{\bf %
p^{\prime }}}W_{{\bf pp^{\prime }}}^{ss{\bf ^{\prime }}}\cos ({\bf p},{\bf %
p^{\prime }}),  \label{17}
\end{equation}
where $s,s^{\prime }=1,2$. The solution of Eqs. (\ref{15}) is

\begin{eqnarray}
A_{1} &=&\frac{\nu _{1}[\omega _{c}^{2}+\nu _{2}^{2}+\eta _{1}\tilde{\nu}%
_{12}\nu _{2}]-\tilde{\nu}_{12}[\tilde{\nu}_{12}\nu _{2}+\eta _{1}(\omega
_{c}^{2}+\tilde{\nu}_{12}^{2})]}{\Delta }  \nonumber \\
B_{1} &=&\frac{\omega _{c}^{2}+\nu _{2}^{2}+\tilde{\nu}_{12}^{2}+\eta _{1}%
\tilde{\nu}_{12}(\nu _{1}+\nu _{2})}{\Delta },  \label{18}
\end{eqnarray}
with the cyclotron frequency $\omega _{c}=|e|H/mc$, $\nu _{1}=\nu
_{1}^{tr}+\nu _{12}$, $\nu _{2}=\nu _{2}^{tr}+\nu _{12}$, and $\Delta =[(\nu
_{1}\nu _{2}-\tilde{\nu}_{12}^{2})^{2}+\omega _{c}^{2}(\omega _{c}^{2}+\nu
_{1}^{2}+\nu _{2}^{2}+2\tilde{\nu}_{12}^{2})]$. In Eq. (\ref{18}) we have $%
\eta _{1}=p^{\prime }/p$, where the momenta $p^{\prime }$ and $p$ are
connected through $(p^{2}-p^{\prime 2})/(2m)=\varepsilon _{2}-\varepsilon
_{1}-2\delta \varepsilon _{{\bf r}}$ from energy conservation. The
expressions for $A_{2}$ and $B_{2}$ follows from the above expressions for $%
A_{1}$ and $B_{1}$, respectively, by changing $1\rightarrow 2$, $%
2\rightarrow 1$ everywhere in Eq. (\ref{18}) and taking $\eta _{2}=p^{\prime
}/p$ where $p^{\prime }$ now satisfies $(p^{2}-p^{\prime
2})/(2m)=\varepsilon _{1}-\varepsilon _{2}+2\delta \varepsilon _{{\bf r}}$.

Furthermore, for the zero-temperature, the coefficients $A_{s}$ and $B_{s}$
are calculated only at the Fermi level. Thus, we have to use $p=p_{F_{1}}$ ($%
p=p_{F_{2}}$) in $A_{1}$,$B_{1}$ ($A_{2}$,$B_{2}$) while $\eta
_{1}=p_{F_{2}}/p_{F_{1}}$ and $\eta _{2}=p_{F_{1}}/p_{F_{2}}$
correspondingly. The induced current density, at zero temperature, is
written as

\begin{equation}
{\bf j}_{{\bf r}}=\frac{e^{2}}{m}\sum_{s}n_{{\bf r}}^{(s)}\left\{ A_{s}{\bf E%
}_{{\bf r}}+B_{s}\frac{e}{mc}[{\bf E}_{{\bf r}}\times {\bf H}]\right\}
\equiv \hat{\sigma}({\bf r}){\bf E}_{{\bf r}},  \label{19}
\end{equation}
where the electron concentration for the $s$-th subband is given by

\begin{equation}
n_{{\bf r}}^{(s)}=n_{s}+(-1)^{s}\rho _{2D}\delta \varepsilon _{{\bf r}}.
\label{20}
\end{equation}
$n_{s}$ is the mean concentration of the $s$-th level and $n=n_{1}+n_{2}$ is
the total average concentration. From Eq. (\ref{19}), the components of the $%
2\times 2$ conductivity tensor are

\begin{equation}
\sigma _{xx}({\bf r})=\frac{e^{2}}{m}\sum_{s}A_{s}n_{{\bf r}%
}^{(s)},\;\;\;\sigma _{yx}({\bf r})=\frac{e^{2}}{m}\omega
_{c}\sum_{s}B_{s}n_{{\bf r}}^{(s)},  \label{21}
\end{equation}
where $\sigma _{xx}({\bf r})=\sigma _{yy}({\bf r})$ and $\sigma _{yx}({\bf r}%
)=-\sigma _{xy}({\bf r})$. Note, that the conductivity dependence on ${\bf r}
$ comes not only due to the spatial variations of $n_{{\bf r}}^{(s)}$ but
also because $p_{F_{1}\text{ }}$and $p_{F_{2}}$ depends on ${\bf r}$.

\section{Transport properties}

In this section, we study the peculiarities of the steady-state
magnetotransport caused by unscreened fluctuations of the QW's energy
levels. In the model of random potentials, the transition probability can be
written as

\begin{equation}
W_{{\bf pp^{\prime }}}^{ss^{\prime }}=\frac{2\pi w}{\hbar L^{3}}\sum_{q}\exp
[-(Q/q_{0})^{2}]|\left\langle s^{\prime }|e^{iqz}|s\right\rangle |^{2}\delta
(E_{{\bf p^{\prime }},s^{\prime },{\bf r}}-E_{{\bf p},s,{\bf r}}),
\label{22}
\end{equation}
where $w\exp (-Q^{2}/q_{0}^{2})$ is the Gaussian correlation function for
the random potential with characteristic scale $q_{0}^{-1}$ (note that $%
q_{0}\gg \ell ^{-1}$), ${\bf Q}=[({\bf p^{\prime }}-{\bf p})/\hbar ,q]$ is a
3D wave vector, and $L$ is the normalization length. Summing over $q$ in Eq.
(\ref{22}) leads to

\begin{equation}
K_{ss^{\prime }}=\frac{1}{L}\sum_{q}e^{-(q/q_{0})^{2}}|\left\langle
s^{\prime }|e^{iqz}|s\right\rangle |^{2}=\frac{q_{0}}{2\sqrt{\pi }}\int
dz\int dz^{\prime }\varphi _{s^{\prime }z}\varphi _{sz}\varphi _{s^{\prime
}z^{\prime }}\varphi _{sz^{\prime }}e^{-[q_{0}(z-z^{\prime })/2]^{2}}
\label{23}
\end{equation}
and the transition probability transforms into

\begin{equation}
W_{{\bf pp^{\prime }}}^{ss^{\prime }}=\frac{2\pi w}{\hbar L^{2}}\exp \{-[(%
{\bf p^{\prime }}-{\bf p})/\hbar q_{0}]^{2}\}K_{ss^{\prime }}\delta (E_{{\bf %
p^{\prime }},s^{\prime },{\bf r}}-E_{{\bf p},s,{\bf r}}),  \label{24}
\end{equation}
where the $\delta $-function depends on ${\bf r}$ only for intersubband
transitions, $s\neq s^{\prime }$. Notice that $K_{12}=K_{21}$.

The functions $K_{ss^{\prime }}$ are calculated below for the cases of
short-range ($q_{0}^{-1}\ll d$) and small-angle ($\hbar q_{0}\ll p_{F{s}}$)
scattering. We consider three models of confinement potential:{\it \ }%
hard-wall QWs (see Sec. IIA),{\it \ }parabolic QWs\cite{8,9} with effective
width $\ell _{\bot }=\sqrt{\hbar /m\omega _{\bot }}$, $\omega _{\bot }$ is
the characteristic frequency of the potential, and{\it \ }stepped QWs\cite
{10} with quite different well widths $d_{n}\ll d_{w}$. The coefficients $%
K_{ss^{\prime }}$ are shown in Table I.

Consider next the relaxation frequencies for these cases. For short-range
scattering, straightforward calculations of Eqs. (\ref{16}), (\ref{17}), and
(\ref{24}) give

\begin{equation}
\nu _{1}^{tr}=(mw/\hbar ^{3})K_{11},\;\;\nu _{2}^{tr}=(mw/\hbar
^{3})K_{22},\;\;\nu _{12}=(mw/\hbar ^{3})K_{12},  \label{25}
\end{equation}
and $\tilde{\nu}_{12}=0$. For small-angle scattering, we obtain

\begin{equation}
\nu _{1,2}^{tr}\approx \frac{mwq_{0}^{3}}{8\sqrt{\pi }p_{F_{1,2}}^{3}}%
K_{11},\;\text{and \ }\nu _{12}\approx \tilde{\nu}_{12}\approx \frac{mwq_{0}%
}{2\sqrt{\pi }\hbar ^{2}\sqrt{p_{F_{1}}p_{F_{2}}}}%
K_{12}e^{-(p_{F_{1}}-p_{F_{2}})^{2}/\hbar ^{2}q_{0}^{2}},  \label{26}
\end{equation}
where we assumed that $(\hbar q_{0}/p_{F_{s}})^{2}\ll 1$ and $p_{F_{s}}$ is
determined from $E_{p{\bf r}}^{(s)}=E_{F}$. We point out that, in these
conditions for small-angle scattering, we have $\nu _{12}\ll \nu _{s}^{tr}$
due to the exponentially small factor. Indeed, for instance, in the case of
the parabolic QW, $\nu _{12}/\nu _{1}^{tr}=\sqrt{p_{F_{1}}/p_{F_{2}}}%
(p_{F_{1}}\ell _{\bot }/\hbar )^{2}\exp [-(p_{F_{1}}-p_{F_{2}})^{2}/\hbar
^{2}q_{0}^{2}]$. For the maximum ratio $p_{F_{1}}/p_{F_{2}}$, we estimate $%
\nu _{12}/\nu _{1}^{tr}=4\sqrt{2}\exp [-(p_{F_{1}}-p_{F_{2}})^{2}/\hbar
^{2}q_{0}^{2}]$. So, if $\exp [-(p_{F_{1}}-p_{F_{2}})^{2}/\hbar
^{2}q_{0}^{2}]\ll 10^{-1}$, we obtain $\nu _{12}/\nu _{s}^{tr}\ll 1$ for
hard-wall and parabolic QWs. Similar condition holds for the stepped QW. For
small-angle scattering, we neglect further very small contributions related
to $\tilde{\nu}_{12}$ and $\nu _{12}$. Then, as it is seen in the Table I
and Eq. (\ref{26}), the transport coefficients do not depend essentially on
the specific QW for small-angle scattering.

In order to analyze the influence of non-uniform contributions, we have to
average the local conductivity tensor, Eqs. (\ref{21}), written in the form $%
\sigma _{\mu \nu }({\bf r})=\left\langle \sigma _{\mu \nu }\right\rangle
+\sum_{{\bf k}}(\delta \sigma _{{\bf k}})_{\mu \nu }\exp (i{\bf k\cdot r})$,
where $\delta \hat{\sigma}_{{\bf k}}$ is the non-uniform contribution and $%
\left\langle \sigma _{\mu \nu }\right\rangle $ is the conductivity tensor
for an ideal QW with two occupied subbands given by

\begin{equation}
\left\langle \sigma _{\mu \nu }\right\rangle =\frac{e^{2}}{m}\sum_{s}\frac{%
n_{s}}{\nu _{s}^{2}+\omega _{c}^{2}}\left\{ 
\begin{array}{ll}
\nu _{s} & \mu =x,\;\nu =x, \\ 
\omega _{c} & \mu =y,\;\nu =x
\end{array}
\right. .  \label{27}
\end{equation}
Here we assume that intersubband frequencies are negligible. Following Ref. 
\cite{12}, we obtain, up to the second order in the large-scale disorder,
the effective conductivity tensor, from $\left\langle {\bf j}_{{\bf r}%
}\right\rangle =\hat{\sigma}^{eff}\left\langle {\bf E}_{{\bf r}%
}\right\rangle $, as

\begin{equation}
\sigma _{\mu \nu }^{eff}=\left\langle \sigma _{\mu \nu }\right\rangle -\sum_{%
{\bf k}}\frac{\sum_{\alpha \beta }\left\langle (\delta \sigma _{{\bf -k}%
})_{\mu \alpha }k_{\alpha }k_{\beta }(\delta \sigma _{{\bf k}})_{\beta \nu
}\right\rangle }{\sum_{\alpha \beta }k_{\alpha }\sigma _{\alpha \beta
}k_{\beta }},  \label{28}
\end{equation}
where $\left\langle ...\right\rangle $ denotes the average over the sample
area, {\it i.e}., for distances $r\gg \ell $, and $\left\langle \sum_{{\bf k}%
}(\delta \sigma _{{\bf k}})_{\mu \nu }\exp (i{\bf kr})\right\rangle =0$ \cite
{12}. The average is performed using Fourier transform of the Gaussian
correlation function, Eq. (\ref{9}), $\langle \delta \varepsilon _{{\bf k}%
}\delta \varepsilon _{{\bf k^{\prime }}}\rangle =\delta _{{\bf k+k^{\prime }}%
,0}(\pi \overline{\delta \varepsilon }^{2}\ell _{c}^{2}/L^{2})\exp [-(k\ell
_{c}/2)^{2}]$ with $k=\sqrt{k_{x}^{2}+k_{y}^{2}}$.

\subsection{Short-range scattering}

Before explicit calculation of the local magnetoconductivity tensor from the
general Eq. (\ref{21}), we present now standard expressions for
magnetotransport coefficients of a QW with two occupied subbands. Using the
conductivity tensor, given by Eq. (\ref{27}), we obtain the magnetoresistance

\begin{equation}
\Delta \rho _{\bot }=\frac{n_{1}n_{2}\omega _{c}^{2}(\nu _{1}-\nu _{2})^{2}}{%
\nu _{1}\nu _{2}[(n_{1}+n_{2})^{2}\omega _{c}^{2}+(n_{1}\nu _{2}+n_{2}\nu
_{1})^{2}]},  \label{29}
\end{equation}
and the Hall coefficient

\begin{equation}
R=\frac{1}{ec}\frac{n_{1}(\nu _{2}^{2}+\omega _{c}^{2})+n_{2}(\nu
_{1}^{2}+\omega _{c}^{2})}{[n_{1}^{2}(\nu _{2}^{2}+\omega
_{c}^{2})+n_{2}^{2}(\nu _{1}^{2}+\omega _{c}^{2})+2n_{1}n_{2}(\nu _{2}\nu
_{1}+\omega _{c}^{2})]}.  \label{30}
\end{equation}
For strong ($\omega _{c}\tau \gg 1$) and weak ($\omega _{c}\tau \ll 1$)
magnetic fields, the Hall coefficient do not depend on $H$ leading to $%
R(\infty )\approx 1/ec(n_{1}+n_{2})$ and $R(H=0)\approx (n_{1}\nu
_{2}^{2}+n_{2}\nu _{1}^{2})/[ec(n_{1}\nu _{2}+n_{2}\nu _{1})^{2}]$.

For short-range scattering, all frequencies do not depend on $p_{F_{1,2}}$
and as a consequence, the coefficients $A_{s}$ and $B_{s}$ are independent
of ${\bf r}$. The non-uniform contribution to the conductivity tensor takes
the form

\begin{equation}
\delta \sigma _{\mu \nu }({\bf k})=\frac{e^{2}\rho _{2D}}{m}\sum_{s}\frac{%
(-1)^{s}\delta \varepsilon _{{\bf k}}}{\nu _{s}^{2}+\omega _{c}^{2}}\left\{ 
\begin{array}{ll}
\nu _{s} & \mu =x,y\;\nu =x,y \\ 
\omega _{c} & \mu =y,\;\nu =x
\end{array}
\right. .  \label{31}
\end{equation}
After substituting Eq. (\ref{31}) into Eq. (\ref{28}) and performing the
averaging process, we obtain the components of the effective conductivity
tensor as

\begin{equation}
\sigma_{xx}^{eff}=\left\langle\sigma_{xx}\right\rangle-\frac{e^{4} \overline{%
\delta n}^{2}}{2m^{2}\left\langle\sigma _{xx}\right\rangle} \left(\frac{1}{%
\omega _{c}^{2}+\nu _{1}^{2}}+\frac{1}{\omega _{c}^{2}+ \nu_{2}^{2}}-\frac{%
2(\omega _{c}^{2}+\nu _{1}\nu _{2})}{(\omega _{c}^{2}
+\nu_{1}\nu_{2})^{2}+\omega_{c}^{2}(\nu_{2}-\nu_{1})^{2} }\right) ,
\label{32}
\end{equation}

\begin{equation}
\sigma _{yx}^{_{eff}}=\left\langle \sigma _{yx}\right\rangle -\frac{e^{4}%
\overline{\delta n}^{2}}{m^{2}\left\langle \sigma _{xx}\right\rangle }\omega
_{c}\left( \frac{1}{\omega _{c}^{2}+\nu _{1}^{2}}-\frac{1}{\omega
_{c}^{2}+\nu _{2}^{2}}\right) \left( \frac{\nu _{1}}{\omega _{c}^{2}+\nu
_{1}^{2}}-\frac{\nu _{2}}{\omega _{c}^{2}+\nu _{2}^{2}}\right) ,  \label{33}
\end{equation}
where $\overline{\delta n}=\rho _{2D}\overline{\delta \varepsilon }$ gives
the concentration fluctuation due to large-scale inhomogeneities.

Using Eqs. (\ref{32}) and (\ref{33}), the effective magnetoresistance $%
\Delta \rho _{eff}$, and the Hall coefficient $R_{eff}$, can be easily
obtained. In Fig. 2, we plot $\Delta \rho _{eff}$ as a function of the
magnetic field for $n_{2}/n_{1}=0.5$. The solid and dot-dashed curves are $%
10 $ times enlarged and correspond to parabolic QWs where $\nu _{2}/\nu
_{1}=5/6 $ and $\overline{\delta n}/n_{1}=0.2$ and $\overline{\delta n}%
/n_{1}=0$ respectively, The dotted and dashed curves correspond to stepped
QWs for $\overline{\delta n}/n_{1}=0.2$ and $\overline{\delta n}/n_{1}=0$,
respectively, with $d_{n}\approx 10$ \AA , $d_{w}\approx 350$ \AA , $%
U_{0}\approx 0.1$ eV, which leads to $\kappa \approx 10^{6}\;$cm$^{-1}$, $%
\varepsilon _{2}-\varepsilon _{1}\approx 21.5$ meV, and $n_{1}+n_{2}\approx
2.0\times 10^{12}\;$cm$^{-2}$. It follows that $\nu _{2}/\nu _{1}=1/2.1$.
Notice that $\Delta \rho _{eff}\equiv 0$ as well as $\Delta \rho _{\bot
}\equiv 0$ for the hard-wall QW because $\nu _{2}/\nu _{1}=1$ for
short-range scattering. Then magnetoresistance is absent for hard-wall QWs,
while a positive magnetoresistance appears for the parabolic QW, which
becomes larger by a factor of $15$ in the case of the stepped QW.

In Fig. 3, we depict $(R_{eff}-R_{0})/R_{0}$, where $R_{0}=1/ecn$ with $%
n=(n_{1}+n_{2})$ using the same curves and conditions as in Fig. 2.
Similarly to Fig. 2, the solid and dot-dashed curves represent $10\times
(R_{eff}-R_{0})/R_{0}$ for the parabolic QW. In the case of the hard-wall
QW, $R_{eff}\equiv R_{0}$ and $R\equiv R_{0}$ because $\nu _{2}/\nu _{1}=1$.
Hence for hard-wall QW $(R_{eff}-R_{0})/R_{0}=0$, while an explicit
dependence of $H$ is observed for parabolic and stepped QWs. In addition,
from Figs. 2,3 it follows that unscreened effect of the large-scale
inhomogeneities of a QW parameters is manifested through stronger
magnetoresistance at $\omega _{c}/\nu _{1}<1$ and weaker at $\omega _{c}/\nu
_{1}>1$. While it leads to smaller Hall coefficient for $\omega _{c}/\nu
_{1}<0.5$, for $\omega _{c}/\nu _{1}\agt2$ effect of such large-scale
inhomogeneities on Hall coefficient becomes negligible. Notice that in
experiment of Ref. \cite{8} it was observed for magnetic fields, $B_{z}<0.2$
T, practically independent of $H_{z}$ magnetoresistance in GaAs-based
parabolic QW with two occupied subbands: which leads to $\nu _{1}$ very
close to $\nu _{2}$ that in turn can be valid only when scattering is
short-range.

\subsection{Small-angle scattering}

For small-angle scattering, large-scale fluctuations of the electron density
and scattering frequencies play an important role. As we have seen,
scattering frequencies, given by Eq. (\ref{26}), depend on fluctuations of
the Fermi momenta. Further, we can neglect contributions coming from $\nu
_{12}$ and $\tilde{\nu}_{12}$. Using Eqs. (\ref{18}), (\ref{21}), (\ref{26}%
), and (\ref{28}), the components of the effective conductivity tensor, for
small-angle scattering, can be written as

\begin{equation}
\sigma _{xx}^{eff}/\overline{\sigma }_{0}=\overline{\sigma _{xx}}-\frac{4%
\overline{\delta n}^{2}}{n^{2}\overline{\sigma _{xx}}}\left[
F_{x}^{2}(H)-F_{y}^{2}(H)\right] ,  \label{34}
\end{equation}
and

\begin{equation}
\sigma _{yx}^{eff}/\overline{\sigma }_{0}=\overline{\sigma _{yx}}-\frac{8%
\overline{\delta n}^{2}}{n^{2}\overline{\sigma _{xx}}}F_{y}(H)F_{x}(H),
\label{35}
\end{equation}
where

\begin{equation}
\overline{\sigma _{xx}}=\left\langle \sigma _{xx}\right\rangle /\overline{%
\sigma }_{0}=2\sqrt{2}\sum_{s=1,2}\frac{(n_{s}/n_{2D})^{1/2}}{1+\omega
_{c}^{2}/\nu _{s}^{2}}\left\{ \left( \frac{n_{s}}{n}\right) ^{2}+\frac{%
\overline{\delta n}^{2}}{n^{2}}\left[ -\frac{19}{16}+\left( \frac{7}{4}-%
\frac{3\omega _{c}^{2}/\nu _{s}^{2}}{1+\omega _{c}^{2}/\nu _{s}^{2}}\right)
^{2}\right] \right\} ,  \label{36}
\end{equation}

\begin{equation}
\overline{\sigma _{yx}}=\left\langle \sigma _{yx}\right\rangle /\overline{%
\sigma }_{0}=2\sqrt{2}\sum_{s=1,2}\frac{(\omega _{c}/\nu _{s})(n_{s}/n)^{1/2}%
}{1+\omega _{c}^{2}/\nu _{s}^{2}}\left\{ \left( \frac{n_{s}}{n}\right) ^{2}+%
\frac{\overline{\delta n}^{2}}{n^{2}}\left[ -\frac{1}{4}+\left( \frac{5}{2}-%
\frac{3\omega _{c}^{2}/\nu _{s}^{2}}{1+\omega _{c}^{2}/\nu _{s}^{2}}\right)
^{2}\right] \right\} ,  \label{37}
\end{equation}
and

\begin{equation}
F_{x}(H)=\sum_{s=1,2}(-1)^{s}\frac{(n_{s}/n)^{3/2}}{1+\omega _{c}^{2}/\nu
_{s}^{2}}\left( \frac{5}{2}-\frac{3\omega _{c}^{2}/\nu _{s}^{2}}{1+\omega
_{c}^{2}/\nu _{s}^{2}}\right) ,  \label{38}
\end{equation}

\begin{equation}
F_{y}(H)=\sum_{s=1,2}(-1)^{s}\frac{(\omega _{c}/\nu _{s})(n_{s}/n)^{3/2}}{%
1+\omega _{c}^{2}/\nu _{s}^{2}}\left( 4-\frac{3\omega _{c}^{2}/\nu _{s}^{2}}{%
1+\omega _{c}^{2}/\nu _{s}^{2}}\right) .  \label{39}
\end{equation}
Here $\nu _{s}=mwq_{0}^{4}/16\pi \overline{p}_{F_{s}}^{3}$, where $\overline{%
p}_{F_{s}}$ is independent of ${\bf r}$ and is determined by $n_{s}$, $%
\overline{\sigma }_{0}=e^{2}n/m\nu _{0}$, and a typical scattering frequency 
$\nu _{0}$ is determined as $\nu _{0}^{-2/3}=(\nu _{1}^{-2/3}+\nu
_{2}^{-2/3})/2$. Notice that $\nu _{s}$ is related to $n_{s}$, while $\nu
_{0}$ corresponds to $n/2$. In Figs. 4 and 5 we present our calculated
effective conductivity tensor, given by Eqs. (\ref{34})-(\ref{39}),
magnetoresistance $\Delta \rho _{eff}$ and the Hall coefficient, $R_{eff}$.

In Fig. 4, we show $\Delta \rho _{eff}$ for $n_{2}/n_{1}=0.5$ as a function
of $\omega _{c}/\nu _{1}$. The solid and dashed curves represent $\Delta
\rho _{eff}$ for small-angle scattering for $\overline{\delta n}/n=1/8$ and $%
\overline{\delta n}/n=0$, respectively. and $\nu _{2}/\nu _{1}=8$. For
comparison, we plot the dotted curve for short-range scattering in the
stepped QW as shown in Fig. 2. We observe in Fig. 4 that, for small-angle
scattering, large-scale inhomogeneities strongly modifies the effective
magnetoresistance in the case of two occupied subbands. Its behavior can be
essentially different from the situation of short-range scattering. In this
case, as it can be seen in Fig. 2, the magnetoresistance increases before
reaching a plateau at $\omega _{c}/\nu _{1}\gtrsim 1$. On the contrary, for
small-angle scattering, this behavior is observed at $\omega _{c}/\nu
_{1}\gtrsim 8$. As it is seen in Fig. 4, there is a much larger region of
classical magnetic fields in which the magnetoresistance increases in case
of small-angle scattering than for short-range scattering. Furthermore, in
contrast with the short-range scattering case, we see that large-scale
unscreened inhomogeneities lead to substantially smaller values of
magnetoresistance at relatively weak magnetic fields, $\omega _{c}/\nu
_{1}\lesssim 3$, while contributes to enhance the magnetoresistance at
larger magnetic fields, $\omega _{c}/\nu _{1}\gtrsim 4$. Notice that
according experimental results of Ref. \cite{11}, where all data are given
for $n_{2}/n_{1}\approx 1/2$, the rather small magnetoresistance was
attributed to small-angle scattering, which is confirmed by our calculated $%
\Delta \rho _{eff}$ given by the solid curve in Fig. 4.

Using the same parameters of Fig. 4, $(R_{eff}-R_{0})/R_{0}$ is depicted in
Fig. 5. We now observe that large-scale unscreened inhomogeneities lead to a
substantially weaker dependence of $R_{eff}$ on $H$ in agreement with the
experimental results of Ref. \cite{11}. In conclusion, our theoretical
results describe qualitatively well the experimental ones.\cite{11} We point
out that the results, in the case of small-angle scattering, are independent
of QW confinement model.

\subsection{Transverse field effect on the conductivity}

Now we consider the modulation of the effective conductivity under
transverse voltage which causes intersubband redistribution of the electron
population. From Eq. (\ref{32}), for zero magnetic field, the effective
conductivity, for short-range scattering, can be written as

\begin{equation}
\sigma ^{eff}/\overline{\sigma }=\left( 1+\frac{\Delta n\Delta \nu }{n%
\overline{\nu }}\right) \left\{ 1-2\frac{\overline{\delta n}^{2}\Delta \nu
^{2}}{n^{2}\overline{\nu }^{2}(1+\Delta n\Delta \nu /n\overline{\nu })^{2}}%
\right\} ,  \label{40}
\end{equation}
where $\overline{\sigma }=e^{2}n\overline{\nu }/m\nu _{1}\nu _{2}$, $%
\overline{\nu }=(\nu _{1}+\nu _{2})/2$, $\Delta \nu =(\nu _{2}-\nu _{1})/2$, 
$\Delta n=(n_{1}-n_{2})=\rho _{2D}(\varepsilon _{2}-\varepsilon _{1})$ with $%
n_{1}=(n+\Delta n)/2$ and $n_{2}=(n-\Delta n)/2$. For hard-wall QWs, because 
$\Delta \nu =0$, it follows that $\sigma ^{eff}$ is independent of $\Delta n$%
. In Fig. 6, we present $\Delta \sigma ^{eff}=\sigma ^{eff}/\overline{\sigma 
}-1$ as a function of $\Delta n/n$. The dotted and dashed curves correspond
to stepped QWs with $\Delta \nu /\overline{\nu }\approx -0.36$, for the same
parameters used in Fig. 2, and $\overline{\delta n}/n=1/8$ and $\overline{%
\delta n}/n=0$, respectively. The thin-solid curve corresponds to a
parabolic QW, with $\Delta \nu /\overline{\nu }=-1/11$, for $\overline{%
\delta n}/n=1/8$ which practically coincides with that for $\overline{\delta
n}/n=0$. We observe that all curves, plotted in Fig. 6, for short-range
scattering, have a negative slope.

For small-angle scattering, the effective conductivity, given by Eq. (\ref
{34}), transforms, for $H=0$, into

\begin{eqnarray}
\sigma ^{eff}/\overline{\sigma }_{0} &=&\frac{1}{2}\left\{ (1+\frac{\Delta n%
}{n})^{5/2}+(1-\frac{\Delta n}{n})^{5/2}+\frac{15\overline{\delta n}^{2}}{%
2n^{2}}\left[ (1+\frac{\Delta n}{n})^{1/2}+(1-\frac{\Delta n}{n})^{1/2}%
\right] \right\}  \nonumber \\
\ast &&\times \left( 1-\frac{25\overline{\delta n}^{2}}{2n^{2}}\frac{\left[
(1+\frac{\Delta n}{n})^{3/2}-(1-\frac{\Delta n}{n})^{3/2}\right] ^{2}}{%
\left\{ (1+\frac{\Delta n}{n})^{5/2}+(1-\frac{\Delta n}{n})^{5/2}+\frac{15%
\overline{\delta n}^{2}}{2n^{2}}\left[ (1+\frac{\Delta n}{n})^{1/2}+(1-\frac{%
\Delta n}{n})^{1/2}\right] \right\} ^{2}}\right) .  \label{41}
\end{eqnarray}
In Fig. 6, we show the results of $\Delta \sigma ^{eff}=\sigma ^{eff}/%
\overline{\sigma }_{0}-1$ as a function of $\Delta n/n$, calculated from Eq.
(\ref{41}) and represented by solid and dot-dot-dashed curves, which
correspond to $\overline{\delta n}/n=1/8$ and $\overline{\delta n}/n=0$,
respectively. We see that in contrast with the short-range scattering, now
the curves have a positive slope. We do not discuss here the relation
between $\Delta n/n$ and the applied voltage. According to, for instance
Refs. \cite{9}, \cite{11} and electro-optical measurements( for references
see Ref. \cite{14}), the population redistribution can be essential already
for rather small voltages. More detailed behavior can be achieved by
performing self-consistent calculations for band structures with diagrams
shown in Figs. 1({\it b-d}).

\section{Concluding remarks}

We have shown that non-screened variations of the intersubband energy modify
essentially the transport properties of quantum wells when two subbands are
occupied. Relative changes in the transport coefficients are determined by
the degree of the random redistribution of the electron density and by the
extent of the asymmetry of scattering processes that, for short-range
scattering, are characterized by dimensionless parameters $\overline{\delta n%
}/n$ and $\Delta \nu /\overline{\nu }$, respectively. Our model is founded
on some assumptions, which we discuss below, and more detailed numerical
calculations, which take into account peculiarities of the band structure
and scattering mechanisms, are necessary for a full description of the
electron transport. However, our results demonstrate that the kinetic
coefficients change when we move from single to double subband occupancy in
samples of similar quality.

Along with magnetotransport characteristics (magnetoresistance, the Hall
coefficient, etc.) and the field effect (modulation of effective
conductivity by transverse voltage), other transport phenomena like
Shubnikov-de Haas oscillations and cyclotron resonance are influenced by
non-screened variations of subbands. For instance, essential broadening of
the peak of intersubband transitions of a QW, due to unscreened large-scale
inhomogeneities,\cite{13} takes place also in the case of two occupied
subbands. The effect of such inhomogeneities on intersubband optical
transitions also requires special study.

We discuss now the model assumptions. We have used the approximation of
smooth inhomogeneities in several points: {\it i)} dependence of QW energy
levels upon the in-plane coordinate ${\bf r}$; {\it ii)} we have assumed
that characteristic scale of smooth inhomogeneities $\ell \gg a_{B}$, and
the screened potential, given by Eq. (\ref{7}), results only on the average
of variations of the energy levels; {\it iii)} the condition that the
transport scattering length $\bar{v}\bar{\tau}\ll \ell $ mainly determines
the regime of applicability of the treatment; {\it iv)} the approximation of
small amplitudes for smooth inhomogeneities, which leads to conditions $%
\overline{\delta n}/n_{1,2}\ll 1$, reduces quantitatively the effects, but
it allows to use theory applied to weakly large-scale inhomogeneous media 
\cite{12}. In addition, our calculations are based on simple QW spectra
models instead of detailed self-consistent calculations of the band
structure. However, we believe that the model assumptions, which are
commonly used, do not change the overall behavior and numerical estimates of
the magnetotransport coefficients. We hope our work can stimulate further
experimental study of the transport properties of nonideal QWs with
multisubband occupancy in order to verify the effects of large-scale,
unscreened inhomogeneities.

\acknowledgements

This work was supported by grants Nos. 95/0789-3 and 98/10192-2 from Funda\c{%
c}\~{a}o de Amparo \`{a} Pesquisa de S\~{a}o Paulo (FAPESP). O. G. B and N.
S. are grateful to Conselho Nacional de Desenvolvimento Cient\'{i}fico e
Tecnol\'{o}gico (CNPq) for research fellowship.\newline

\newpage

TABLE I. Coefficients $K_{ss^{\prime }}$, which appear in the transition
probability given by Eq. (\ref{24}). The columns labeled {\it I} and {\it II}
stand for the short-range and small-angle scatterings, respectively, and for
({\it b}) hard-wall, ({\it c}) parabolic and ({\it d}) stepped QWs depicted
in Fig. 1. $\hbar \kappa =\sqrt{2m(\varepsilon _{21}-\varepsilon _{w})}$
with $\varepsilon _{w}=2(\pi \hbar /d_{w})^{2}/m$, $\kappa d_{n}\ll 1$, $%
\kappa d_{w}\gg 1$, and $q_{0}d_{w}\ll 1$ and $a\approx 4.6\times 10^{-3}$.

\bigskip

\begin{tabular}{|c|c|c|c|c|c|c|}
\hline
& {\it I (b)} & {\it I (c)} & {\it I (d)} & {\it II (b)} & {\it II (c)} & 
{\it II (d)} \\ \hline
$K_{11}$ & $3/(2d)$ & $1/(\sqrt{2\pi }\ell _{\bot })$ & $\kappa /2$ & $%
q_{0}/(2\sqrt{\pi })$ & $q_{0}/(2\sqrt{\pi })$ & $q_{0}/(2\sqrt{\pi })$ \\ 
\hline
$K_{22}$ & $3/(2d)$ & $3/(4\sqrt{2\pi }\ell _{\bot })$ & $3/(2d_{w})$ & $%
q_{0}/(2\sqrt{\pi })$ & $q_{0}/(2\sqrt{\pi })$ & $q_{0}/(2\sqrt{\pi })$ \\ 
\hline
$K_{12}$ & $1/d$ & $1/(2\sqrt{2\pi }\ell _{\bot })$ & $4\pi
^{2}/[d_{w}(\kappa d_{w})^{2}]$ & $aq_{0}(q_{0}d)^{2}$ & $q_{0}(q_{0}\ell
_{\bot })^{2}/(8\sqrt{\pi })$ & $32\pi ^{3/2}q_{0}^{3}/[\kappa ^{2}(\kappa
d_{w})^{3}]$ \\ \hline
\end{tabular}

\vspace{0.5cm}

\vspace{0.5cm}

\begin{center}
{\bf FIGURE CAPTIONS}
\end{center}

\medskip

FIG. 1. ({\it a}) Schematic views of the spatial variations of two occupied
energy levels in the QW along the $x$ direction with (solid curve) and
without (dotted curve) screening. Band diagrams for ({\it b}) hard-wall, (%
{\it c}) parabolic and ({\it d}) stepped QWs with non-ideal heterointerfaces.

\vspace{0.7cm}

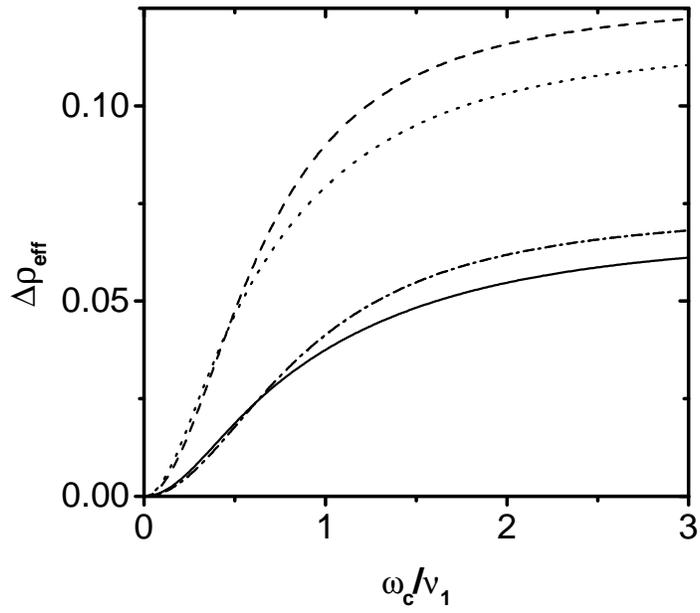
FIG. 2. Magnetoresistance $\Delta \rho _{eff}$ as a function of the magnetic
field for short-range scattering. The solid and dash-dotted curves plot $%
10\times \Delta \rho _{eff}$ in case of parabolic QWs, $\nu _{2}/\nu
_{1}=5/6 $, with $\overline{\delta n}/n_{1}=0.2$ and $\overline{\delta n}%
/n_{1}=0$, respectively. The dotted and dashed curves correspond to stepped
QWs, $\nu _{2}/\nu _{1}=1/2.1$, with $\overline{\delta n}/n_{1}=0.2$ and $%
\overline{\delta n}/n_{1}=0$, respectively. The ratio between average
subband populations $n_{2}/n_{1}=1/2$ and $\overline{\delta n}$ is the
typical value of the concentration fluctuations in both occupied subbands.

\vspace{0.7cm}

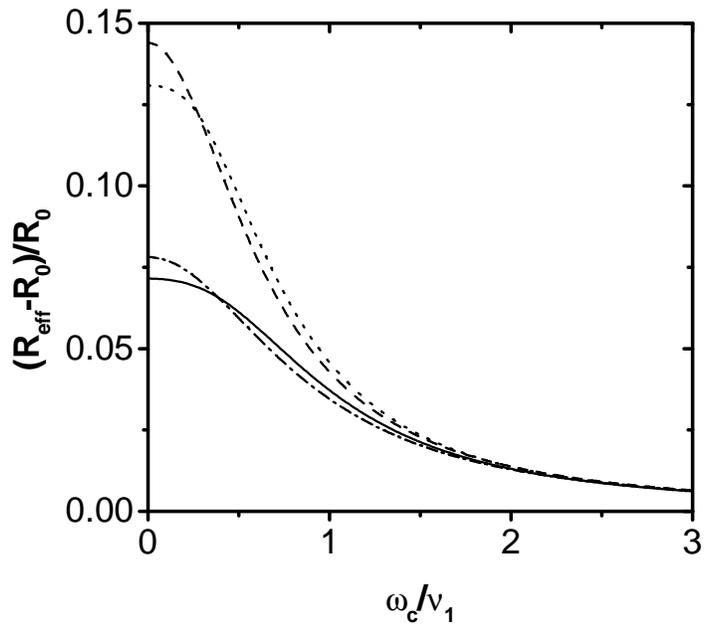
FIG. 3. Hall coefficient $R_{eff}$ as function of magnetic field for
short-range scattering, where $R_{0}=1/ecn$, $n=n_{1}+n_{2}$. The curves
represent the same cases as pertinent curves in Fig. 2.

\vspace{0.7cm}

FIG. 4. Magnetoresistance $\Delta \rho _{eff}$ as function of magnetic field
for small-angle scattering. The solid and dashed curves correspond to $%
n_{2}/n_{1}=1/2$ with $\overline{\delta n}/n=1/8$ and $\overline{\delta n}%
/n=0$, respectively. For comparison with the $\Delta \rho _{eff}$ behavior
for short-range scattering, the dotted curve is the same as in Fig. 2.

\vspace{0.7cm}

FIG 5. Hall coefficient $R_{eff}$ as function of magnetic field for
small-angle scattering. The solid, dashed, and dotted curves correspond to
pertinent curves in Fig. 4.

\vspace{0.7cm}

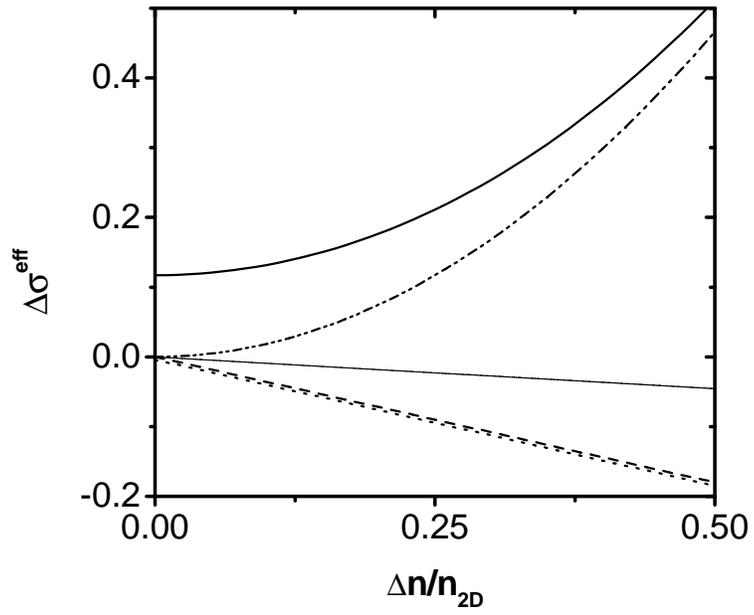
FIG. 6. Conductivity $\Delta \sigma ^{eff}$, for zero magnetic field, as a
function of $\Delta n=n_{1}-n_{2}$ showing the effect of intersubband
population redistribution due to transverse electric field. The solid and
dot-dot-dashed curves, with a positive slope, are for small-angle scattering
and $\overline{\delta n}/n=1/8$ and $\overline{\delta n}/n=0$, respectively.
The thin solid (parabolic QW with $\overline{\delta n}/n=0$) dotted (stepped
QW with $\overline{\delta n}/n=1/8)$, and dashed curves (stepped QW with $%
\overline{\delta n}/n=0)$ are for short-range scattering; other parameters
are the same as for the stepped QW in Fig. 2.


\begin{references}
\bibitem{1}  E. D. Siggia, P. C. Kwok, Phys. Rev. B {\bf 2}, 1024 (1970).

\bibitem{2}  S. Mori, T. Ando, Phys. Rev. B {\bf 19}, 6433 (1979).

\bibitem{3}  H. L. Stormer, A. C. Gossard, W. Wiegmann, Solid State Commun. 
{\bf 41}, 707 (1982).

\bibitem{4}  H. van Houten, J. G. Williamson, and M. E. I. Broekaart, Phys.
Rev. B {\bf 37}, 2756 (1988).

\bibitem{5}  R. Fletcher, E. Zaremba, M. D'Iorio, C. T. Foxon, and J. J.
Harris, Phys. Rev. B {\bf 41}, 10649 (1990); D. R. Leadley, R. Fletcher, R.
J. Nicolas, F. Tao, C. T. Foxon, J. J. Harris, Phys. Rev B {\bf 46}, 12439
(1992).

\bibitem{6}  G.-Q. Hai, N. Studart, and F. M. Peeters, Phys. Rev. B {\bf 52}%
, 8363 (1995); G.-Q. Hai, N. Studart, G. E. Marques, F. M. Peeters, P. M.
Koenraad, Physica E {\bf 2}, 222 (1998).

\bibitem{7}  V. Piazza, P. Casarini, S. De Franceschi, M. Lazzarino, F.
Beltram, C. Jacoboni, A. Bosacchi, and S. Franchi, Phys. Rev. B {\bf 57},
10017 (1998).

\bibitem{8}  K. Ensslin, A. Wixforth, S. Sundaram, P. F. Hopkins, J. H.
English, and A.C. Gossard, Phys. Rev. B {\bf 47}, 1366 (1993).

\bibitem{9}  G. Salis, B. Graf, K. Ensslin, K. Campman, K. Maranowski and
A.C. Gossard, Phys. Rev. Lett. {\bf 79}, 5106 (1997).

\bibitem{10}  P. F. Yuh, K. L. Wang, J. Appl. Phys. {\bf 65}, 4377 (1989);
V. Berger, Semicond Sci. Technol. {\bf 9}, 1493 (1994); H. Li, Z. Wang, J.
Liang, B. Xu, J. Xu, Q. Gong, C. Jiang, F. Liu, and W. Zhou, J. Appl. Phys. 
{\bf 82}, 6107 (1997).

\bibitem{11}  S. A. Studenikin, A. V. Chaplik, I. A. Papaev, G. Salis, K.
Ensslin, K. Maranowski and A. C. Gossard, Semicond Sci. Technol. {\bf 14},
604 (1999).

\bibitem{12}  C. Herring, Journ. Appl. Phys. {\bf 31}, 1939 (1960).

\bibitem{13}  F. T. Vasko, J. P. Sun, G. I. Haddad, and V. V. Mitin, J.
Appl. Phys., April, (2000).

\bibitem{14}  F. T. Vasko and A. Kuznetsov, {\em Electronic States and
Optical Transitions in Semiconductor Heterostructures}, Springer-Verlag,
Berlin (1998).
\end{references}
\end{document}